\documentclass[a4paper]{article}

\usepackage[english]{babel}
\usepackage[utf8]{inputenc}
\usepackage[colorinlistoftodos, color=green!40, prependcaption]{todonotes}
\usepackage{amsthm}
\usepackage{mathtools}
\usepackage{physics}
\usepackage{xcolor}
\usepackage{graphicx}
\usepackage[left=23mm,right=13mm,top=35mm,columnsep=15pt]{geometry} 
\usepackage{adjustbox}
\usepackage{placeins}
\usepackage[T1]{fontenc}
\usepackage{lipsum}
\usepackage{csquotes}
\newcommand{\RomanNumeralCaps}[1]
    {\MakeUppercase{\romannumeral #1}}
\usepackage[utf8]{inputenc} 
\usepackage{newunicodechar}
\newunicodechar{​}{}
\DeclareUnicodeCharacter{2009}{\,}

\usepackage{hyperref}

\usepackage{subcaption}

\usepackage{graphicx}
\usepackage{ragged2e}
\usepackage{sidecap}
\usepackage{amsmath}
\usepackage{bigints}
\usepackage[T1]{fontenc}
\usepackage[utf8]{inputenc}
\usepackage{authblk}

\DeclareUnicodeCharacter{2212}{\ensuremath{-}}

\begin{document}

\title{\huge Decoupling Optical and Thermal Responses: Thermo-optical Nonlinearities Unlock MHz Transmission Modulation in Dielectric Metasurfaces }
 
\author[]{Omer Can Karaman}
\author[]{Gopal Narmada Naidu}
\author[]{Alan R. Bowman}
\author[]{Elif Nur Dayi}
\author[]{Giulia Tagliabue\thanks{Correspondence email address:~giulia.tagliabue@epfl.ch}}
\affil[]{Laboratory of Nanoscience for Energy Technologies (LNET), STI, École Polytechnique Fédérale de Lausanne, 1015 Lausanne, Switzerland}

\maketitle
\begin{abstract}


Thermo-optical nonlinearities (TONL) in metasurfaces enable dynamic control of optical properties like transmission, reflection, and absorption through external stimuli such as laser irradiation or temperature.  As slow thermal dynamics of extended systems are expected to limit modulation speeds ultimately, research has primarily focused on steady-state effects. In this study, we investigate photo-driven TONL in amorphous silicon (a-Si) metasurfaces both under steady-state and, most importantly, dynamic conditions (50 kHz modulation) using a 488 nm continuous-wave pump laser. First, we show that a non-monotonic change in the steady-state transmission occurs at wavelengths longer than the electric-dipole resonance (800 nm). In particular,  at 815 nm transmission first decreases by 30\% and then increases by 30\% as the laser intensity is raised to 5 mW/$\mu$m$^2$. Next, we demonstrate that TONL decouple the thermal and optical characteristic times, the latter being up to 7 times shorter in the tested conditions (i.e $\tau_{opt}=$0.5 $\mu s$ vs $\tau_{th}=$3.5 $\mu s$). Most remarkably, we experimentally demonstrate that combining  these two effects enables optical modulation at twice the speed (100 kHz) of the excitation laser modulation. We finally show how to achieve all-optical transmission modulation at MHz speeds with large amplitudes (85\%). Overall, these results show that photo-driven TONL produce large and fully reversible transmission modulation in dielectric metasurfaces with fast and adjustable speeds. Therefore, they open completely new opportunities toward exploiting TONL in dynamically reconfigurable systems, from optical switching to wavefront manipulation.

\end{abstract}

\section{Introduction} \label{sec:Intro}

Tunable metasurfaces are of great importance in the fields of optics and photonics due to their ability to manipulate electromagnetic waves across a wide range of frequencies\cite{gu_reconfigurable_2023,shalaginov_design_2020}. Unlike traditional static optical elements, tunable metasurfaces offer flexibility in real-time control of properties such as phase, amplitude, and polarization of light \cite{barton_wavefront_2021,holsteen_temporal_2019,wu_tunable_2019,zhang_tunable_2017,wang_all-optical_2024}. The mechanisms of tunability in active metasurfaces generally involve external stimuli such as electrical, optical, mechanical, or thermal inputs that alter the material's optical properties. Electro-refractive tuning is primarily achieved by integrating materials with high electro-optic coefficients, such as liquid crystals\cite{chung_tunable_2020}, \RomanNumeralCaps{3}-\RomanNumeralCaps{5} semiconductors \cite{wu_dynamic_2019} or lithium niobate (LN), \cite{weigand_enhanced_2021}, into the metasurface design. Optical tuning, instead, is typically achieved by exploiting nonlinear optical effects, such as Kerr effect\cite{shcherbakov_ultrafast_2017}, or photo-induced processes, such as hot carriers generation in plasmonic metals\cite{zhu_ultralow-power_2014,karaman_ultrafast_2024,kiani_transport_2023} or free carrier generation in semiconductors \cite{gu_active_2012}.

Thermo-optical and photo-thermal nonlinearities have been widely explored recently for their potential in tuning optical responses in nanostructures and metasurfaces. Indeed metasurfaces exploiting Mie resonances in high-refractive-index dielectric materials, such as silicon, can confine light efficiently and generate strong field enhancements within the nanoresonators. This is particularly valuable in enabling nonlinear optical phenomena in silicon, which are too weak in the bulk material​\cite{duh_giant_2020,zograf_all-dielectric_2021,li_nonlinear_2021,yang_nonlinear_2015}. The ability to modulate light dynamically in such systems has implications for various advanced photonic applications, including optical switching\cite{zangeneh_kamali_reversible_2019}, beam steering\cite{horie_high-speed_2018,kim_thermo-optic_2019,rabinovich_two-dimensional_2016}, and polarization manipulation\cite{wang_all-optical_2024,wu_tunable_2019,zhang_tunable_2017}. In recent works exploring photo-thermal and thermo-optical nonlinearities, the focus has predominantly been on steady-state nonlinear behavior in silicon-based metasurfaces and nanostructures \cite{zograf_all-dielectric_2021,che_ultrasensitive_2024}. These studies have demonstrated impressive modulation of the optical properties through both photo-thermal effects, where absorbed light induces heating, and thermo-optical effects, where temperature changes modify the refractive index. For example, silicon nanostructures have been shown to modulate scattering and transmission intensities under continuous-wave (CW) excitation, with effective nonlinear indices $n_2$ orders of magnitude higher than bulk silicon (i.e, $n=n_0+n_2I$ where $I$ is the excitation intensity)​\cite{duh_giant_2020,zhang_anapole_2020,li_nonlinear_2021}. Recently, the studies introduced quasi-bound states in the continuum, q-BIC, in silicon metasurfaces to enable thermo-optical bistability, and optimize hysteresis width and switching power, suggesting a path for energy-efficient optical computing and non-reciprocity using silicon-based metasurfaces \cite{cotrufo_passive_2024,barulin_thermo-optical_nodate}.

However, while steady-state studies have achieved significant breakthroughs, transient modulation through thermo-optical and photo-thermal nonlinearities has been thought to be inherently limited by the thermal response time of the system. This is particularly evident when photo-thermal effects, which involve localized heating and changes in the optical properties, are used for dynamic modulation​\cite{duh_giant_2020}. Most metasurfaces and nanostructures investigated for their thermo-optical responses, such as those supporting high-Q resonances, are bound by the timescales required for heat diffusion, limiting their application in scenarios requiring fast modulation speeds \cite{huang_transient_2022,nishida_all-optical_2023}.

\begin{figure} [t!]
    \centering
   \includegraphics[]
   {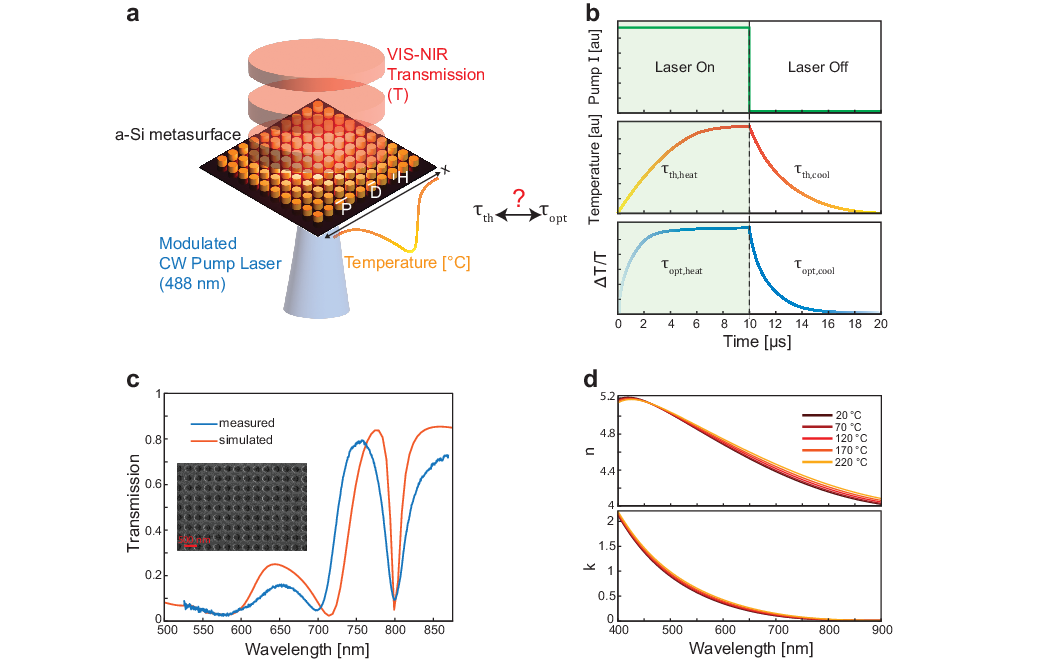}
    \caption{\justifying{ \textbf{Optical properties of the proposed a-Si metasurface.} \textbf{a} The schematic illustration of a-Si metasurface on Fused Silica substrate (P, D, and H denote the periodicity, diameter, and height of the disks, respectively.). \textbf{b} The schematic illustration of the modulated CW laser signal (upper), the temperature evolution in time (middle) and differential transient transmission (lower), $\Delta T/T$, of the metasurface. \textbf{c} The transmission spectra of the fabricated (blue) and simulated (orange) metasurfaces. The inset shows the SEM image of the fabricated metasurface with P of 380 nm, D of 290 nm, and H of 100 nm. \textbf{d} The measured real part, n, and imaginary part, k, of the refractive indices of a-Si at 20 $^o$C, 70 $^o$C, 120 $^o$C, 170 $^o$C, and 220 $^o$C.  }}
    \label{fig1}
\end{figure}
In this study, we investigate the photo-thermo-optical response and nonlinear dynamics of amorphous silicon (a-Si) metasurfaces (\textbf{Figure \ref{fig1}a}), demonstrating their unique capability to decouple thermal and optical characteristic timescale as well as their potential for photothermally-driven transmission modulation, both in terms of amplitude and speed. We designed and fabricated high-Q electric dipole resonant metasurfaces and characterized their transmission spectrum, $T(\lambda)$, where $\lambda$ is the wavelength, under both steady-state and modulated (50 kHz) photo-excitation using a CW 488 nm laser with variable intensity and beam size. Under steady-state irradiation, we demonstrate large, nonlinear, and fully reversible changes in $T(\lambda)$, with both the sign (i.e., increase or decrease) and intensity dependence determined by the wavelength position relative to resonance. On resonance (i.e. 800 nm), a monotonic increase of $T$ from 0.09 to 0.51 ($\approx 470\%$ experimental, $\approx 1360\%$ theoretical) is obtained under 5 mW/$\mu m^2$ irradiation. Strikingly, at 815 nm (red-shifted from resonance), a non-monotonic response is reported, with $T$ first decreasing from to 0.22 (-31\% at 1.5 mW/$\mu m^2$) and then increasing to 0.41 (+29\% at 5 mW/$\mu m^2$). For modulated irradiation, we further show that thermo-optical non-linearities (TONL) uniquely decouple the system’s thermal response time ($\tau_{th}$) from its optical response time ($\tau_{opt}$) ((\textbf{Fig. \ref{fig1}b})). For example, with a beam diameter of 3 $\mu$m, exciting approximately 7 by 7 a-Si nanoresonators, the heating dynamics exhibit $\tau_{th,heat}$ (the time interval between 10\% and 90\% of the temperature change) $\approx$3.5 $\mu$s, while the transmission, $T(t)$, response time decreases with increasing excitation intensity, reaching $\tau_{opt,heat}$ (the time interval between 10\% and 90\% of the transmission change) $\approx$0.5 $\mu$s at 3.3 $mW/\mu m^2$. Most remarkably, we demonstrate that the combination of non-monotonic $T(t)$ response and decoupled $\tau_{th}/\tau_{opt}$ can lead to a doubling of the optical modulation frequency (100 kHz) with respect to the photothermal modulation frequency (50 kHz) while preserving a large modulation amplitude (30\% experimental, 85\% theoretical). Finally, we discuss how $\tau_{th}/\tau_{opt}$ decoupling can also enable photo-thermo-optical modulation speeds exceeding 1 MHz. Our experimental findings are in excellent agreement with a Multiphysics COMSOL model, which incorporates newly measured, temperature-dependent refractive index (n) and extinction coefficient (k) data for a-Si over 20–220°C. Overall, these results highlight that photo-driven TONL in dielectric metasurfaces open completely new opportunities toward dynamically reconfigurable systems, including optical switching devices \cite{zangeneh_kamali_reversible_2019} and wavefront manipulation components, such as beam steering \cite{kim_thermo-optic_2019,rabinovich_two-dimensional_2016} and metalenses \cite{archetti_thermally_2022,shalaginov_reconfigurable_2021}.

\section{Results} \label{sec:Results}
\begin{figure} [ht!]
    \centering  
   \includegraphics[]
   {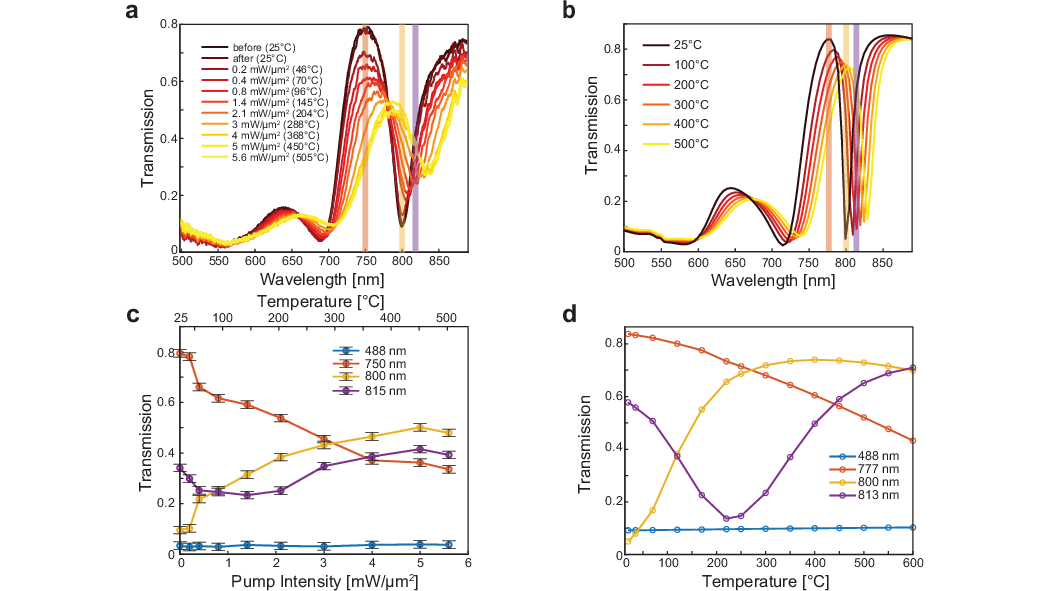}
    \caption{\justifying{ \textbf{The measured and simulated TONL in steady-state.} \textbf{a} The measured transmission spectra of the metasurface under different pump intensities. The legend specifies the temperatures measured by Raman thermometry at the corresponding pump intensities. The semi-transparent regions specify the probe wavelengths in \textbf{c}. \textbf{b} The simulated transmission spectra of the metasurface under different pump intensities. The semi-transparent regions specify the probe wavelengths in \textbf{d}. \textbf{c} The measured pump intensity-dependent transmission of the metasurface at the probe wavelengths of 488 nm (blue), 750 nm (orange), 800 nm (yellow), and 815 nm (violet). Upper x-axis shows the measured temperatures corresponding to the pump intensities. \textbf{d} The simulated pump intensity (temperature)-dependent transmission of the metasurface at the probe wavelengths of 488 nm (blue), 777 nm (orange), 800 nm (yellow), and 813 nm (violet).}}
    \label{fig2}
\end{figure}

\textbf{Optical properties of the a-Si metasurface.} 
We designed and fabricated a-Si metasurfaces consisting of an array of nanodisks with diameters ($D$) ranging from 200 nm to 300 nm,  a height ($H$) of 100 nm and a periodicity ($P$) of 380 nm. The experimental transmission spectrum of the metasurface with nanodisks of 290 nm diameter (\textbf{Fig. \ref{fig1}c}, blue curve) is characterized by two pronounced dips around 695 nm and 800 nm, in good agreement with the simulated results (orange curve). The resonance at 695 nm is attributed to the magnetic dipole (MD) mode, while the resonance at 800 nm corresponds to the electric dipole (ED) mode. We provide detailed field profiles for both the MD and ED resonances in \textbf{Supplementary Fig. 1}, illustrating the field confinement and mode distributions at these resonant wavelengths. The quality factor (Q-factor) of the experimental ED resonance at 800 nm is $\approx$ 50, reflecting the metasurface's ability to achieve strong light confinement and offer good possibility for optical modulation. This is made possible by the high quality of the fabricated nanodisk array (see inset of \textbf{Fig. \ref{fig1}c}).  While there is a close agreement between the simulated and measured spectra, slight deviations are likely due to small fabrication imperfections or unaccounted material loss in the simulations. Nonetheless, both the experimental and simulated data confirm the strong resonant behavior of the metasurface within the visible to near-infrared spectrum, showcasing its potential for applications in optical modulation.

To further explore the thermo-optical properties of the metasurface, the temperature dependence of the refractive index of an a-Si thin film (100 nm thick) is measured by ellipsometry using an external heater to vary the sample temperature from 20°C to 220°C (\textbf{Fig. \ref{fig1}d}). The measurement shows that across the visible and near-infrared wavelength range, a-Si thermo-optical coefficient is dispersive and can be negative or positive depending on the wavelength (see \textbf{Supplementary Fig. 5}), differently from previously assumed constant values across the spectrum\cite{shcherbakov_ultrafast_2015,fauchet_ultrafast_1989,della_valle_nonlinear_2017}. The real (n) and imaginary (k) parts of the refractive index of a-Si exhibit a more pronounced increase in the spectral window between 650 nm and 850 nm as the temperature increases. This temperature dependence reflects the thermo-optic effect in a-Si, which can be exploited to dynamically tune the optical properties of the metasurface under varying thermal conditions. Notably, the measured thermo-optical coefficient, derived from the variation of the refractive index (n) with temperature, exhibits a maximum ($\approx$ 4x10$^{-4}$ K$^{-1}$) at $\approx$ 800 nm, coinciding with the ED resonance (see \textbf{Supplementary Fig. 5a}). This alignment between the peak thermo-optical coefficient and the ED resonance highlights the metasurface as an ideal platform for exploring thermo-optical effects. The strong field confinement at the ED resonance further amplifies the temperature-induced refractive index changes, making it particularly suited for dynamic tuning of optical properties via thermal modulation. Note that our measurements reveal that the real part of the thermo-optical coefficient of a-Si is 0 at $\approx$ 480 nm, which makes this spectral region important for applications requiring high stability against temperature changes\cite{chang_integrated_2022,kodheli_satellite_2021}. 

\textbf{TONL in steady-state optical response }
To explore the impact of TONL on the steady-state transmission spectrum of a-Si metasurfaces, we perform a series of transmission measurements under varying photoexcitation intensities (\textbf{Figure \ref{fig2}}). A 488 nm continuous wave (CW) laser is used as \textit{pump} to photoexcite/heat the metasurface, which has an absorption of $\approx$ 0.55 at this wavelength. Simultaneously, a low-intensity, CW white light source is utilized as the \textit{probe} to measure the transmission spectra across the visible/near-infrared spectrum, allowing us to quantify the change of the metasurface's optical transmission in real-time at selected probe wavelengths. Further details of the experimental setup, including a diagram, are provided in the Methods section and \textbf{Supplementary Note 2}. Taking advantage of the pump CW laser,  we also use calibrated Raman thermometry to measure the temperature of the metasurface in-situ during the experiments (comprehensive details are provided in the Methods section).


\textbf{Figure \ref{fig2}a} displays the measured transmission spectra of the metasurface as a function of pump laser intensity (3 $\mu$m beam diameter), with the corresponding temperatures ranging from 20°C to 500°C. As the pump intensity increases, the transmission dips associated with MD at around 695 nm and ED at around 800 nm red shift and become less pronounced. Notably, the ED resonance experiences a larger shift ($\approx$40 nm) than the MD resonance ($\approx$22 nm), which can be attributed to two main reasons. First, the thermo-optical coefficient of a-Si is higher in the spectral region of the ED resonance than at the MD resonance position, leading to a greater refractive index change with temperature near 800 nm. Second, the ED mode exhibits a higher field enhancement and Q-factor than the MD, resulting in stronger light-matter interactions and greater sensitivity to temperature-induced refractive index changes. This enhanced field confinement can induce larger shifts in the ED resonance as the metasurface heats up. When cooling the sample (after exposure to the maximum pump intensity), the transmission spectrum returns to its initial state, indicating that the applied pump intensity can reversibly tune the metasurface's optical properties without permanent modifications of the materials, such as laser-induced crystallization or oxidation of a-Si\cite{nayak_femtosecond-laser-induced-crystallization_2007,berzins_laser-induced_2020,marcins_crystallization_2011}. To model the temperature-dependent response of the metasurface, we first extrapolated the a-Si experimental refractive index data beyond 220°C using an approach that we validated against available literature for other materials \cite{nishida_all-optical_2023,hoyland_temperature_2006,vuye_temperature_1993,cherroret_temperature-dependent_2008} (see \textbf{Supplementary Note 4}). Simulations are in excellent agreement with the experimental data, capturing the trends in transmission modulation with increasing laser intensity \textbf{Figure \ref{fig2}b}. 

To assess the existence and magnitude of nonlinear variations in the transmitted signal due to thermo-optical effects, particularly near the ED resonance at 800 nm, we analyze the evolution of transmission at selected wavelengths. The pump intensity-dependent transmission at four probe wavelengths (488 nm, 750 nm, 800 nm, and 815 nm) is plotted in \textbf{Figure \ref{fig2}c}, clearly showing a nonlinear response of the metasurface at wavelengths around the ED resonance. For the 488 nm probe, which is far from resonance, there is minimal change in transmission, indicating that the thermo-optical effect is less pronounced at this wavelength because of the low thermo-optical coefficient of a-Si and the flat spectrum. In contrast, at 750 nm (left of resonance) and 800 nm (on resonance), there is a significant negative and positive nonlinear change in the transmission as the pump intensity increases, respectively. The nonlinearity is particularly strong at 800 nm, where the ED resonance occurs, leading to a sharp increase in transmission with increasing pump intensity, reaching up to $\approx$ 470$\%$ $\Delta T/T$, (from 0.09 to 0.51) at $\approx$ 5.6 mW/$\mu$m$^2$ intensity. At 815 nm probe wavelength, we also observe a unique behavior: transmission first decreases from 0.32 to 0.22 (-30\% at $\approx$ 1.5 $mW/\mu m^2$) and then increases, reaching the initial value at $\approx$ 3 mW/$\mu$m$^2$ and then raising up to 0.42 (+30\%)  for pump a pump intensity of $\approx$ 5 mW/$\mu$m$^2$, eventually plateauing at higher intensities. Numerical simulations confirm the trends observed in experiments, with significant transmission modulation occurring near the resonant wavelengths (777 nm, 800 nm, and 813 nm). In comparison, the off-resonance wavelength (488 nm) remains largely unaffected (\textbf{Fig. \ref{fig2}d}). The good agreement between experimental and simulated results demonstrates the accuracy of the material and electromagnetic models in capturing TONL of the metasurface. Based on the calculations, with improvements in the fabrication, up to $\approx$ 1360$\%$ $\Delta T/T$ (from 0.05 to 0.73) is possible thanks to a gigantic TONL at the resonance wavelength of 800 nm. 

\begin{figure} [t!]
    \centering
  
   \includegraphics[]
   {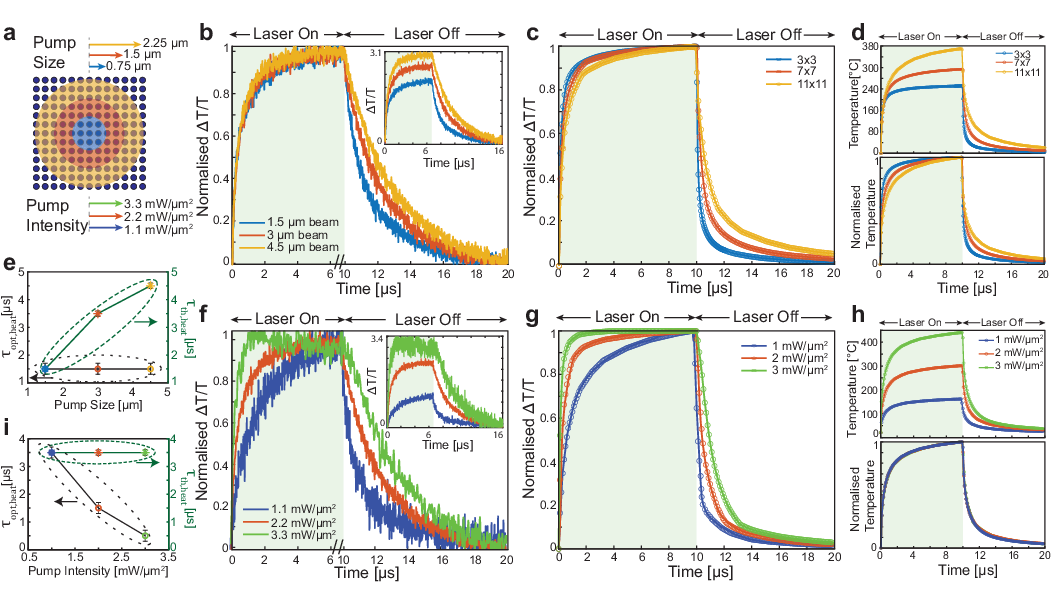}
    \caption{\justifying{ \textbf{The measured and simulated transient TONL at resonance (800 nm). a} The schematic illustration of the metasurface with the pump beams of 0.75 (blue), 1.5 (red), and 2.25 (yellow)$\mu$m in radius used in panel \textbf{b} and the pump intensities of 1.1 (deep blue), 2.2 (red), and 3.3 (green) mW/$\mu$m$^2$ in panel \textbf{f} with the radius of 1.5 $\mu$m. In all measurements, the probe size was 0.75 $\mu$m in radius. \textbf{b} The measured normalised $\Delta T/T$ of the metasurface at 800 nm probe for the 1.5 (blue), 3 (red), and 4.5 (yellow) $\mu$m 488 nm pump beams in diameter at 2.2 mW/$\mu$m$^2$. The inset shows the nonnormalised transmission,  $\Delta T/T$, of the same cases. \textbf{c} The simulated normalised $\Delta T/T$ of the 3 by 3 (blue), 7 by 7 (red), and 11 by 11 (yellow) array of nanodisks at 800 nm probe while pumping at 488 nm at 2 mW/$\mu$m$^2$. \textbf{d} The simulated transient (upper) and normalised transient (lower) temperatures of the 3 by 3 (blue), 7 by 7 (red), and 11 by 11 (yellow) array of nanodisks at 2 mW/$\mu$m$^2$ pump intensity. \textbf{e} The measured $\tau_{opt,heat}$ in \textbf{b} and calculated $\tau_{th,heat}$ in \textbf{d} versus pump diameter. \textbf{f} The measured normalised $\Delta T/T$ of the metasurface at 800 nm probe for 1.1 mW/$\mu$m$^2$ (deep blue), 2.2 mW/$\mu$m$^2$ (red), and 3.3 mW/$\mu$m$^2$ (green) intensity 488 nm 3 $\mu$m in diameter pump beams. The inset shows  $\Delta T/T$ of the same cases. \textbf{g} The simulated normalised $\Delta T/T$ of the 7 by 7 array of nanodisks for the 1 mW/$\mu$m$^2$ (deep blue), 2 mW/$\mu$m$^2$ (red), and 3 mW/$\mu$m$^2$ (green) 488 nm pump intensities at 800 nm probe wavelength. \textbf{h} The simulated transient (upper) and normalised transient (lower) temperatures of the 7 by 7 array of nanodisks for the 1 mW/$\mu$m$^2$ (deep blue), 2 mW/$\mu$m$^2$  (red), and 3 mW/$\mu$m$^2$ (green) 488 nm pump intensities. \textbf{i} The measured $\tau_{opt,heat}$ in \textbf{f} and calculated $\tau_{th,heat}$ in \textbf{h} versus pump intensity.}}
    \label{fig3}
\end{figure}

\begin{figure} [t!]
    \centering
  
   \includegraphics[]
   {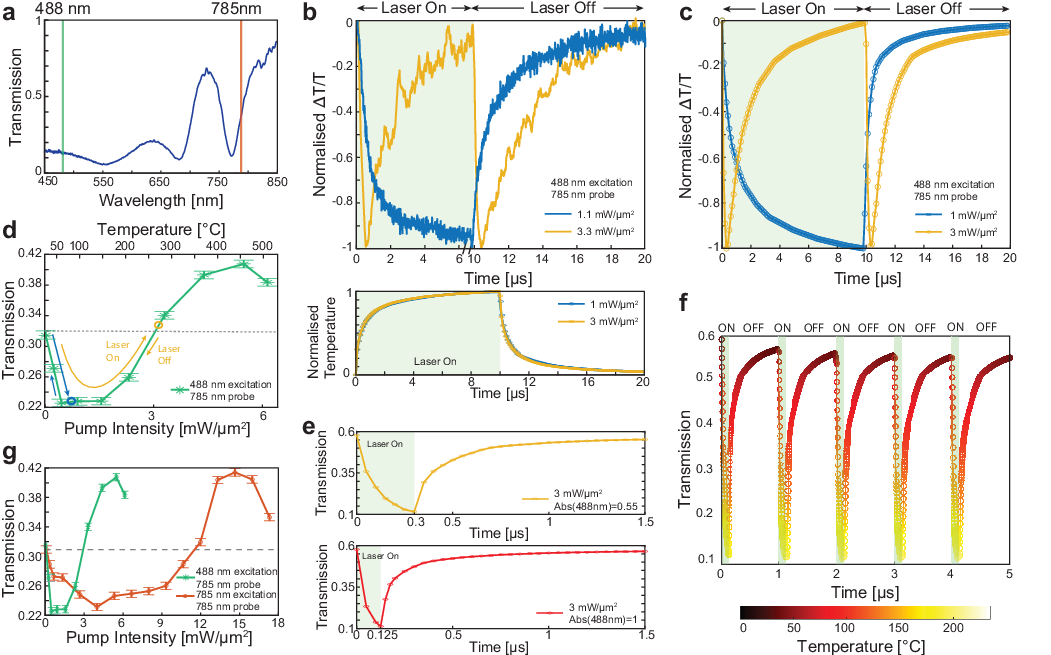}
    \caption{\justifying{ \textbf{The measured and simulated transient TONL at near resonance probing. a} The transmission spectrum of the metasurface with ED resonance at 770 nm and the spectral positions of the pump (green) and probe (orange) wavelengths of 488 nm and 785 nm, respectively. \textbf{b} The measured normalised $\Delta T/T$ (upper) and normalised temperature (lower) of the metasurface, for the 1.1 mW/$\mu$m$^2$ (blue), and 3.3 mW/$\mu$m$^2$ (yellow) 488 nm pump intensities at 785 nm probe wavelength. \textbf{c} The simulated normalised $\Delta T/T$ of the 7 by 7 array of nanodisks for the 1 mW/$\mu$m$^2$ (blue), and 3 mW/$\mu$m$^2$ (yellow) 488 nm pump intensities at 785 nm probe wavelength. \textbf{d} The measured pump intensity-dependent transmission of the metasurface at the pump and probe wavelengths of 488 and 785 nm, respectively. Upper x-axis shows the measured temperatures corresponding to the pump intensities. The blue (yellow) arrow illustrates the gradual change in transmission at 1.1 (3.3) mW/$\mu$m$^2$ in \textbf{b}. \textbf{e} The simulated $\Delta T/T$ at 785 nm (upper) for the laser is turned off at 300 ns at 3 mW/$\mu$m$^2$ pump intensity with 0.55 absorption at 488 nm and (lower) for the laser is turned off at 125 ns at 3 mW/$\mu$m$^2$ pump intensity with unity absorption at 488 nm.  \textbf{f} The simulated transient transmission at 785 nm for the modulated 488 nm laser at 1 MHz with laser-on time equals 125 ns at 3 mW/$\mu$m$^2$ intensity with unity absorption at 488 nm. \textbf{g} The measured pump intensity-dependent transmission of the metasurface at the probe wavelength of 785 nm for the pump wavelengths of 488 nm (blue) and 785 nm (orange).}}
    \label{fig4}
\end{figure}

\textbf{TONL in transient optical response} To investigate the transient TONL in a-Si metasurface, we performed time-resolved measurements and simulations under various pump and probe conditions as demonstrated schematically in \textbf{Figure \ref{fig3}a}. The transient response was measured by probing the metasurface at the resonance wavelength (\textbf{Fig. \ref{fig3}}) and near resonance wavelength (\textbf{Fig. \ref{fig4}}) while pumping it with the 488 nm continuous wave (CW) laser using different beam sizes and intensity levels. The 488 nm CW pump laser was modulated at 50 kHz, with a transient modulation time of about 2 ns, which is much faster than the heating and cooling times of the metasurface. As such, this modulation does not influence the measured transient dynamics, allowing us to capture the thermo-optical response of the system accurately. The details of the $\Delta T/T$ setup can be found in Methods and \textbf{Supplementary Note 2}. The resulting $\Delta T/T$ dynamics provide insight into the interaction between optical heating and the time evolution of the metasurface's refractive index, highlighting the role of transient TONL in modulating light transmission.

\textbf{Figure \ref{fig3}b} shows the experimentally measured normalised $\Delta T/T$ at 800 nm probe wavelength (i.e. at the ED resonance) for three different pump beam radii (0.75 $\mu$m - blue, 1.5 $\mu$m - red, and 2.25 $\mu$m - yellow) but constant pump intensity of 2.2 mW/$\mu$m$^2$. Transmission is instead probed in a central region with a constant diameter of 1.5 $\mu$m, equivalent to approximately 3 by 3 meta-atoms. Increasing the pump beam diameter while keeping the intensity unchanged in the probed region allows us to assess the impact of collective heating effects \cite{naef_light-driven_2023} on transient TONL. In fact, the constant intensity ensures that light absorption and heat dissipation in each meta-atom (i.e. self-heating) are the same in all measurements. On the other hand, illuminating more meta-atoms (larger beam) results in a larger contribution of heat diffusion (i.e. collective heating) onto the final temperature of the meta-atoms in the central region (details of the configuration is in \textbf{Supplementary Note 2}). In agreement with the steady-state results (\textbf{Figure \ref{fig2}c}, yellow curve), for all beam sizes  $\Delta T/T$ increases during the "Laser On" phase and decreases when the laser is turned off. Notably, while the characteristic time of the optical signal in the "Laser On" phase ($\tau_{opt,heat}$), i.e. during heating, is approximately constant, the characteristic time during the "Laser Off" phase ($\tau_{opt,cool}$), i.e. cooling, depends on the beam size, smaller beams leading to faster changes in transmission. Additionally, as shown in the inset of \textbf{Fig. \ref{fig3}b},  the amplitude of $\Delta T/T$ increases with the beam diameter. These observations depend on the elevated temperatures caused by the increase in collective heating effects with the beam diameter. We also performed COMSOL simulations to model the $\Delta T/T$ response of metasurfaces consisting of 3 by 3, 7 by 7, and 11 by 11 arrays of nanodisks and mimic the effect of increasing beam diameters (\textbf{Fig. \ref{fig3}c}) (see Methods for the details of the simulations). These also show that larger arrays (corresponding to larger beams) exhibit slower transient responses due to the more distributed collective heating effects and higher temperatures (\textbf{Fig. \ref{fig3}d}). With the model, we also compute the temperature evolution of each system and the associated thermal characteristic time ($\tau_{th}$) for heating and cooling. We obtain $\tau_{th,heat}$ equal to 1.5, 3.5 and 4.5 $\mu$s, and  $\tau_{th, cool}$ equal to 1.4, 3.2 and 5.6 $\mu$s for 3 by 3, 7 by 7, and 11 by 11 array of nanodisks, respectively. Interestingly, while $\tau_{th,heat}$ becomes slower for larger arrays, $\tau_{opt,heat}$ remains approximately constant and equal to $\approx$ 1.5 $\mu$s, as observed in experiments (\textbf{Fig. \ref{fig3}e}). The increased temperatures at larger beam diameters, caused by collective heating, amplify thermo-optical nonlinear responses in $\Delta T/T$, enhancing transmission dynamics and decoupling them from thermal dynamics. Small differences between the calculated and measured characteristic times of the transmission modulation can arise from the unaccounted cold nanoresonators around the heated region or a minor mismatch in the thermal parameters of the materials. Overall, these results confirm the critical role of beam size in modulating the transient optical response of the metasurface.

The influence of pump intensity on the transient response is shown in \textbf{Figure \ref{fig3}f}, which illustrates the normalised $\Delta T/T$ at low (1.1 mW/$\mu$m$^2$), medium (2.2 mW/$\mu$m$^2$), and high (3.3 mW/$\mu$m$^2$) laser intensity levels. The inset of \textbf{Figure \ref{fig3}f} shows $\Delta T/T$, providing insight into the amplitude changes under different intensity conditions. As expected, higher pump intensities (green curve) result in stronger modulation, i.e. a larger change in transmission as a consequence of elevated temperatures (\textbf{Fig. \ref{fig3}d}), confirming the strong thermo-optical nonlinear effect in the metasurface. 

Contrary to the case of increasing beam size, \textbf{Figure \ref{fig3}h} shows that  $\tau_{th,heat}$  and  $\tau_{th,cool}$ remain constant for different pump intensities, as shown also in previous works \cite{hu_heat_2002,huang_transient_2022}. Specifically, they are equal to 3.5 $\mu$s and 3.2 $\mu$s, respectively. However, in \textbf{Figure \ref{fig3}f} we interestingly observe that $\tau_{opt, heat}$, and $\tau_{opt, cool}$ vary significantly with the pump intensity, $\tau_{opt, heat}$ decreasing from $\approx$ 6 $\mu$s to $\approx$ 0.5 $\mu$s when the intensity increases from 1.1 mW/$\mu$m$^2$ to 3.3 mW/$\mu$m$^2$ (\textbf{Fig. \ref{fig3}i}). This apparent discrepancy is a unique consequence of the TONL on $\Delta T/T$. Although the thermal response of the metasurface (heating and cooling rates) remains unaffected by the pump intensity, TONL accelerate the effect on $\Delta T/T$, resulting in a much faster modulation of the signal at higher intensities. We confirmed the experimental result with numerical simulations, achieving excellent agreement regarding transient times and magnitudes (\textbf{Fig. \ref{fig3}g}). This further highlights how TONL can decouple the thermal response from the transient optical behavior, creating a situation where optical modulation speeds do not directly correspond to the thermal dynamics of the system.

We finally consider the case where we probe the transient response of the metasurface at a wavelength that is red-shifted compared to the ED resonance mode for which a non-monotonic evolution of the transmission signal with incident intensity was observed in steady-state. Because of experimental limitations (available bandpass filters), we used a metasurface comparable to that presented above but with the ED resonance at 770 nm (nanodisk $D$ of 280 nm, $P$ of 380 nm, and $H$ of 100 nm) and probed it at 785 nm (\textbf{Fig. \ref{fig4}a}). \textbf{Figure \ref{fig4}b} (upper), shows the normalized $\Delta T/T$ for 1.1 mW/$\mu$m$^2$ and 3.3 mW/$\mu$m$^2$ pump intensities. A markedly different response from that observed on-resonance (\textbf{Fig. \ref{fig3}f}) is obtained. Most strikingly, there exist clear variations in both the characteristic times and the shape of $\Delta T/T$ with pump intensity, highlighting the profound influence of TONL on the transient response. At 1.1 mW/$\mu$m$^2$ transmission monotonically decreases. Indeed upon photothermal heating the resonance redshifts ($\approx$ 12 nm), eventually matching the probe wavelength, which thus reaches the minimum transmission value. Once the laser is turned off, the resonance shifts back to its initial position resulting in a monotonic increase in the transmission. At 3.3 mW/$\mu$m$^2$, however, as time passes and temperature increases further, the resonance redshifts beyond the probe wavelength, reaching $\approx$ 30 nm shift at steady-state. When the resonance center becomes more red-shifted than the probe wavelength, transmission increases again. We thus observe that when the laser is turned on, transmission shows an extremely fast dip ($\approx$ 300 ns), followed by rapid recovery to the initial transmission value (i.e. zero differential transmission). When the laser is turned off, the system similarly exhibits a fast dip in transmission and a quick recovery. This trend can be better understood by considering \textbf{Figure \ref{fig4}d}, acquired in steady state. During a dynamic modulation with the 488 nm CW laser, the transmission signal evolves in time along the green curve, up to 1.1 mW/$\mu$m$^2$ or 3.3 mW/$\mu$m$^2$ pump intensities, as shown by the blue and yellow arrows, with non-uniform speed due to the non-linear change in temperature. For 3.3 mW/$\mu$m$^2$ pump intensity, the final transmission has the same amplitude as the initial one, as observed by the recovery of $\Delta T/T$ in \textbf{Figure \ref{fig4}d}, yellow curve. Overall, in this particular condition we observe that while the laser is modulated at $50 kHz$, the transmission signal is effectively modulated at twice this speed, i.e. $100 kHz$.  We emphasize that this complex optical behavior is not mirrored by the thermal response, as \textbf{Figure \ref{fig4}b} (lower) shows that the temperature monotonically increases when the laser is on and decreases after it is turned off. This discrepancy between the optical and thermal responses confirms that the system is no longer linear and is governed by highly complex nonlinearities that effectively decouple the transient optical response from the underlying thermal dynamics. In \textbf{Figure \ref{fig4}c}, we present simulated normalized $\Delta T/T$ for a 7 by 7 nanodisk array, resonant at 770 nm, probed at 785 nm as in the experimental conditions. The simulations closely replicate the measured behavior, confirming that the transient optical modulation is governed by both the intensity of the pump laser and the resonant properties of the metasurface. Overall, TONL introduces a decoupling of thermal and optical dynamics and the resonant nature of the metasurface introduces non-monotonic evolutions of the optical signal with temperatures. Together,  they generate unique and non-intuitive behaviors in the metasurface transient optical properties, further reinforcing the importance of TONL in driving advanced optical modulation effects.

To explore the limits of modulation speed in the studied system, we now concentrate on controlling independently the laser-on and laser-off phases. Indeed, although above we demonstrated an effective doubling of the modulation speed, we observe that the transmission recovery time during the laser-off phase increases with increasing incident power. The upper panel in \textbf{Figure \ref{fig4}e} shows the calculated transient transmission when the laser is turned off after 300 ns, instead of 10 $\mu s$, for a pump intensity of 3 mW/$\mu$m$^2$. The initial sharp drop in transmission (from 0.6 to 0.1 corresponds to $\approx$ 85$\%$ change in $T$) follows the behavior reported in \textbf{Figure \ref{fig4}b-c} for the high pump intensity case (yellow curve, laser on phase). However, when the laser is turned off, the cooling dynamic follows the evolution observed for the low pump intensity case (blue curve, laser off phase). This is due to the fact that the cooling process is governed solely by the temperature reached at 300 ns, irrespective of the illumination intensity used. The temperature at the minimum of $\Delta T/T$ is the same for the high and low pump intensity cases, as it corresponds to the shift of the ED resonance to the probe wavelength. However, the heating rate and thus the time it takes to reach this temperature strongly depends on the used intensity. This behavior effectively decouples the heating and cooling rates, enabling faster modulation. The modulation speed, therefore, becomes even faster than what would be expected from traditional thermo-optical switches \cite{horie_high-speed_2018,zangeneh_kamali_electrically_2023}. Finally, we also consider the case in which absorption of the metasurface at 488 nm is unity instead of 0.55 as in our experiments (\textbf{Fig. \ref{fig4}e} lower panel). As just discussed, a higher heating rate further accelerates the optical modulation rate resulting in the minimum transmission occurring at just 125 ns. Moreover, thanks to flash heating, the heated area is more localized resulting in faster cooling and recovery of the transmission when the laser is turned off compared to the case of absorption is equal to 0.55. This unique behavior opens the possibility for operating the metasurface in the MHz range with very high modulation depths (\textbf{Fig. \ref{fig4}f}), making it significantly more efficient than conventional thermo-optical switches, which are typically constrained by slower thermal response times. By leveraging this fast modulation mechanism, we demonstrate that TONL in a-Si metasurfaces has the potential to break the typical limitations of thermal-based modulation, offering much faster switching capabilities.

\textbf{Comparison between the dual and single band operations} When using the 488 nm laser for photothermally heating the metasurface (\textbf{Figure \ref{fig4}a}, green line), the pump wavelength is spectrally far from any resonance, corresponding to dual-band operation. Consequently, absorption at the pump wavelength remains essentially constant with temperature and, as the pump intensity increases, the steady-state temperature of the metasurface increases as well (\textbf{Fig. \ref{fig4}d}).  Therefore, the observed steady-state and dynamic transmission changes are purely due to TONL at the probe wavelength. In contrast, when using a 785 nm pump laser (\textbf{Figure \ref{fig4}a}, orange line), close to the ED resonance of the metasurface, the system operates in a single-band mode, where both thermo-optical and photo-thermal nonlinearities must be considered \cite{nishida_optical_2023,tang_mie-enhanced_2021,tsoulos_self-induced_2020,duh_giant_2020}. As shown in \textbf{Figure \ref{fig4}g}, orange curve, at low pump intensities, transmission decreases as the metasurface heats up but not as much as with 488 nm pumping (green curve) because of the much lower absorption at 785 nm. As the pump intensity increases, the photo-thermal nonlinearity becomes more pronounced, resulting in an enhanced nonlinear response (see \textbf{Supplementary Fig. 6}). Eventually, the evolution of the transmission with pump intensity follows the same shape as with 488 nm excitation but with a non-linear stretching of the x-axis due to the non-linear change in absorbed power. The interaction between the heating effects and the resonance properties of the metasurface thus causes the transmission to behave in a more complex manner compared to the dual-band operation.

\section{Discussion}
Our study reveals novel insights into the transient and steady-state TONL in a-Si metasurfaces, building on existing literature primarily focused on steady-state nonlinear optical effects. The results highlight the unique ability of a-Si metasurfaces to exhibit nonlinear and non-monotonic transmission changes and to decouple the transient optical response from the slower thermal dynamics of the system. These two effects, driven by TONL, eventually enable high-speed transmission modulation while benefiting from the resonant nature of the system to achieve large modulation amplitudes.

The role of collective heating was also explored in our experiments, particularly by varying the pump beam size and intensity. We found that smaller beams induce faster thermal responses due to more localized heating, while larger beams result in slower responses due to the increase of collective heating effects. This behavior is consistent with the predictions of earlier work \cite{kpustovalov_light--heat_2016}, where the spatial distribution of heat plays a significant role in determining the overall optical response of the metasurface. More importantly, the enhanced collective heating effects from larger beams result in higher temperatures, leading to stronger TONL. Consequently, while the thermal response time increases with the size of the irradiated system, the optical response time remains nearly unchanged, allowing for faster modulation times than anticipated. Furthermore, we showed that increasing the pump beam intensity, for a given beam size, can dramatically reduce the optical response characteristic time, while leaving the thermal response characteristic time unaltered. This provides a simple and direct way to control the dynamics of the optical response of the metasurface. 

Finally, we briefly discuss the use of Raman thermometry for in-situ temperature measurements. Raman thermometry is indeed a highly effective method for measuring the temperature of silicon materials \cite{baffou_anti-stokes_2021,xu_raman-based_2020}. Monocrystalline silicon (c-Si) displays a pronounced and distinct Raman peak, which arises from the interaction of light with phonon modes. The temperature of Si nanoparticles can thus be determined based on the linewidth broadening of the Raman peak or its spectral shift as well as from the reduction in the anti-Stokes/Stokes signal ratio \cite{hart_temperature_1970,zograf_resonant_2017}. However, it has been observed that Mie resonances, present in both metals and semiconductors, can have a significant impact on the Raman scattering spectrum. Specifically, they can affect the linewidth of the spectral peaks as well as alter the anti-Stokes/Stokes ratio, with the latter being influenced by the Purcell effect\cite{zograf_stimulated_2020,vikram_photo-thermo-optical_2024}. To mitigate the impact of Mie resonances, we employed a 488 nm pump laser to induce the Raman signal, as the spectrum of the metasurface at this wavelength lacks Mie-like resonances and remains almost flat. Due to the broadened and more chaotic Raman signal in amorphous materials, there have been fewer attempts in the literature to apply Raman thermometry to amorphous silicon\cite{bhusari_temperature-dependent_1993,viera_crystal_2001}. In our approach, we decompose the Raman peaks to isolate the intensity and central positions of each vibrational mode of a-Si. This reveals several vibrational peaks, including a distinct transverse optical phonon mode at 470 cm$^{-1}$ (see \textbf{Supplementary Note 3}). As we showed through the excellent agreement of experiments and simulations, the anti-Stokes/Stokes ratio of this mode can be used to reliably measure the local a-Si temperature during the experiments.

Overall, high-speed all-optical photonic devices are attracting growing interest and our findings provide new guidelines on how to exploit TONL towards rapid modulation. The ability of a-Si metasurfaces to decouple the transient optical response from thermal dynamics presents a new avenue for developing fast, tunable optical modulators, switches, and sensors. To further enhance the transient nonlinearities and modulation rates, one can think of modulating higher Q factor resonances such as q-BIC  \cite{barulin_thermo-optical_nodate} or surface lattice resonances (SLR)\cite{zundel_lattice_2023}, as well as pumping the system at wavelengths with near-unity absorption \cite{xu_radiative_2021,alaee_theory_2017} or utilizing the concept of thermo-optical bistability\cite{cotrufo_passive_2024,barulin_thermo-optical_nodate}. Furthermore, by exploiting nonlinear changes in the absorption rate of the pump beam, the modulation rates can be further enhanced, and complex dynamic responses can be engineered. Moreover, the excellent agreement between our experimental results and COMSOL simulations reinforces the robustness of the TONL mechanism in a-Si metasurfaces, further supporting their potential for high-performance applications in telecommunications and optical computing.

In conclusion, this study provides a comprehensive exploration of both steady-state and transient TONL in a-Si metasurfaces. The novel decoupling of optical and thermal responses opens up new possibilities for achieving ultrafast optical modulation, surpassing the limitations imposed by thermal dynamics in traditional metasurface systems. This work represents a significant step forward in understanding and harnessing TONL for advanced photonic applications.

\section{Methods}
\textbf{Sample preparation.}
The metasurface samples were fabricated on a 550 µm thick fused silica substrate. A 100 nm thick layer of amorphous silicon (a-Si) was deposited onto the substrate using plasma-enhanced chemical vapor deposition (PECVD) at 550°C. This deposition technique ensures a uniform a-Si layer with well-controlled thickness, which is critical for the metasurface's optical performance. Electron beam lithography (EBL-Raith EBPG5000+) was employed to pattern the a-Si nanodisks. A 120 nm thick ZEP 520A resist was spin-coated onto the a-Si layer and subsequently patterned using EBL (100 kV and 200 $\mu C/cm^2$ beam) to define the nanodisk structures. After developing the resist, the patterned ZEP resist served as a mask for the etching process.
The a-Si nanodisks were then etched into the substrate using argon ion beam etching (Veeco Nexus IBE350) (170V and 175mA beam). The back surface of the chips was also etched by IBE to remove the deposited a-Si at the back surface of the chips. After the etching process, the remaining ZEP resist was removed by soaking the samples in acetone. To ensure the complete removal of any residual organic material, the samples were subjected to a final low-power microwave plasma cleaning (Tepla-300) (500W and 400ml/min O$_2$), which further cleaned the surface without affecting the nanodisk structures.
\\
\textbf{Setups.} 
\textit{Steady-state measurements (Fig.\ref{fig1}b, Fig.\ref{fig2}, and Fig.\ref{fig4}d):}
The experimental setup used for spectral, Raman, and $\Delta T/T$ measurements is a custom-built system designed to enable precise control and visualization of the metasurface sample. The setup includes a microscopy section to visually inspect the sample, which utilizes a high numerical aperture (NA 0.8) 100x objective (Nikon-LU Plan ELWD) and a CMOS camera (Thorlabs-	
CS165MU). The halogen lamp (OSL2IR-Thorlabs) is used for the transmission measurements. The system is coupled with an Andor Shamrock 750 spectrometer body equipped with an iDus 420 CCD camera for spectral measurements. The system is equipped with a 488 nm OBIS LX laser, which can be digitally modulated up to 150 MHz. This laser serves as the primary pump source for many of the experiments, including Raman and $\Delta T/T$ measurements. Additionally, a 785 nm laser is integrated into the setup for non-degenerate (single-band) measurements, allowing us to explore the metasurface's behavior under different excitation wavelengths. For Raman and steady-state transmission measurements, the 488 nm pump laser is filtered using a series of cut-off filters to remove any unwanted pump light. The spectrometer is configured with a high line number grating to increase the spectral resolution, ensuring that even subtle shifts in the Raman spectra are captured with high precision.It is worth noting that $\Delta T/T$ is defined as $(T(\lambda, intensity)-T_0(\lambda))/T_0(\lambda)$ where $T_0(\lambda)$ is the unperturbed room temperature transmission.

\textit{Transient measurements (Fig.\ref{fig3} and Fig.\ref{fig4}a):} $\Delta T/T$ measurements are conducted using a commercial time-correlated single photon counting (TCSPC) system (PicoQuant Hydraharp 400) coupled with a silicon single-photon avalanche diode (Si SPAD- Micro Photon Devices PDM Series) detector, providing a time resolution of approximately 20 ps. (NA 0.8) 100x objective is used in these measurements. In these measurements, the 488 nm laser was modulated at 50 kHz using a square wave trigger signal (\% 50 duty cycle) (TTi TG320), which serves as a synchronization signal for the TCSPC module. To selectively probe specific spectral regions of transmitted white light, narrow bandwidth ($\approx$ 3 nm in FWHM) bandpass filters were employed, isolating the wavelengths of interest for detailed analysis. It is worth noting that $\Delta T/T$ is defined as $(T(\lambda, time)-T_0(\lambda))/T_0(\lambda)$. 
\\
\textbf{COMSOL modelling of the TONL.}
All simulations were performed by using the Finite Element Method (FEM) based commercial software COMSOL Multiphysics (COMSOL Multiphysics 6.2). 

\textit{Transient photo-thermal heating simulations (Fig.\ref{fig3}d and Fig.\ref{fig3}e):} The simulated structures were modeled in close agreement with the experimental geometric parameters  (H=100 nm, D=290 nm, and P=380 nm).  A 3D model was built with 3 by 3, 7 by 7, and 11 by 11 arrays of a-Si nanopillars supported on SiO$_2$ to mimic the area irradiated with the pump laser beam as we changed the beam diameter from 1 to 3 $\mu m$. Each structure was placed in a 400 nm thick air layer of refractive index 1, lying on the surface of a 400 nm thick SiO$_2$ background medium layer, with both dimensions along the vertical direction. The lateral dimensions of the simulated domain were varied with the size of the array simulated. The simulation domain was truncated by exploiting 300 nm thick Infinite Element Domains (IEDs) in all of the spatial directions. The nanopillar’s optical properties were described by using the temperature-dependent refractive indices we measured by ellipsometry. A monochromatic plane wave at normal incidence from above was utilized to excite the metasurface. The wavelength of the input light was set to the wavelength corresponding to the pump laser. Laser power was varied in the simulations to match the experimental conditions. The laser illumination was switched on for 10 $\mu s$ and subsequently switched off to study the heating and cooling dynamics of the illuminated metasurface region. Maxwell’s equations were then solved in the transient frequency domain for the total field at the excitation wavelength. The resistive losses could then be evaluated from the model through the volume integral in the subdomain of concern. With the resistive losses calculated from electromagnetic simulation, the resultant temperature dynamics were simulated with a 3D heat transfer model. The nanoparticle array and SiO$_2$ substrate were set as heat sources, with heat dissipated according to the integration of resistive losses. The ambient temperature was set at 20°C and the average temperatures of the simulated particles are considered in the study. We also considered the influence of temperature increase on the heat conductivity and electric permittivity of the materials.

\textit{Metasurface unit cell simulations at elevated temperatures (Fig.\ref{fig2}b):} The metasurface unit cell simulation (Electromagnetic Waves, Frequency Domain) at different temperatures was done by using the measured n and k dataset at different temperatures. Due to the limitation of ellipsometry, we extrapolated the n and k at higher than 220$^0$C (see \textbf{Supplementary Note 4}).

We also mapped the transient temperatures to the perturbed transmission using the abovementioned simulations and obtained $\Delta T/T$ simulations for the specific cases in the study. 
\\
\textbf{Raman thermometry.}
We employed Raman spectroscopy to measure the local temperature of the metasurface, adapting a method previously used for c-Si \cite{duh_giant_2020,hart_temperature_1970,tsu_temperature_1982,zhang_anapole_2020} to a-Si. Unlike c-Si, a-Si exhibits a more complex and broadened Raman spectrum due to the amorphous nature of its phonon modes. The most pronounced phonon mode in a-Si appears at approximately 470 cm$^{-1}$, corresponding to the transverse optical (TO) phonon mode. To determine the local temperature of the metasurface, we fitted Gaussian functions to the measured Raman spectrum to accurately identify the center and amplitude of the 470 cm$^{-1}$ TO mode. The amplitude of this mode was then used to calculate the temperature. Specifically, we applied the same temperature calculation model that utilizes the intensity ratio of the Stokes and anti-Stokes Raman peaks. The equation used is:
\begin{equation}
    \frac{I_A}{I_S} = \exp\left(\frac{-\hbar \omega_p}{k_B T}\right)
\end{equation}
where $I_A$ and $I_S$ are the intensities of Anti-Stokes and Stokes Raman signals, respectively, $k_B$ is the Boltzmann constant, $\hbar$ is the Planck constant, $\omega_p$ is the frequency of the phonon mode, and $T$ is the temperature.

\section*{Data Availability Statement} \label{sec:data}
All the data supporting the findings of this study are presented in the Results section and Supplementary Information are available from the corresponding authors upon reasonable request.


\section*{Acknowledgements} \label{sec:acknowledgements}
GNN and GT acknowledge the support of the Swiss National Science Foundation (Starting Grant 211695). END and GT acknowledge the support of the Swiss National Science Foundation (Eccellenza Grant 194181) and the STI Discovery Grant (EPFL). ARB acknowledges the support of SNSF Eccellenza Grant 194181 and SNSF Swiss Postdoctoral Fellowship TMPFP2$\_$217040. The authors also acknowledge the support of the Center of MicroNanoTechnology (CMi) at EPFL.
\section*{Author contributions} \label{sec:contributions}
G.T supervised all aspects of the project. O.C.K. prepared the sample and performed the experiments and simulations. G.N.N. built the transient simulation model. A.R.B. and E.N.D. played an active role in the discussions and building the setup. G.T. and O.C.K. wrote the paper, with input from all the other authors.
\section*{Competing interests} \label{sec:comp}
The Authors declare no competing interests.


%
\newpage
\section*{Figure Captions} \label{sec:figures}

\noindent \textbf{Optical properties of the proposed a-Si metasurface.} \textbf{a} The schematic illustration of a-Si metasurface on Fused Silica substrate (P, D, and H denote the periodicity, diameter, and height of the disks, respectively.). \textbf{b} The schematic illustration of the modulated CW laser signal(upper), the temperature evolution in time(middle) and differential transient transmission(lower), $\Delta T/T$, of the metasurface. \textbf{c} The transmission spectra of the fabricated (blue) and simulated (orange) metasurfaces. The inset shows the SEM image of the fabricated metasurface with P of 380 nm, D of 290 nm, and H of 100 nm. \textbf{d} The measured real part, n, and imaginary part, k, of the refractive indices of a-Si at 20 $^o$C, 70 $^o$C, 120 $^o$C, 170 $^o$C, and 220 $^o$C. 
\\

\noindent \textbf{The measured and simulated TONL in steady-state.} \textbf{a} The measured transmission spectra of the metasurface under different pump intensities. The legend specifies the temperatures measured by Raman thermometry at the corresponding pump intensities. The semi-transparent regions specify the probe wavelengths in \textbf{c}. \textbf{b} The simulated transmission spectra of the metasurface under different pump intensities. The semi-transparent regions specify the probe wavelengths in \textbf{d}. \textbf{c} The measured pump intensity-dependent transmission of the metasurface at the probe wavelengths of 488 nm (blue), 750 nm (orange), 800 nm (yellow), and 815 nm (violet). Upper x-axis shows the measured temperatures corresponding to the pump intensities. \textbf{d} The simulated pump intensity (temperature)-dependent transmission of the metasurface at the probe wavelengths of 488 nm (blue), 777 nm (orange), 800 nm (yellow), and 813 nm (violet).
\\

\noindent  \textbf{The measured and simulated transient TONL at resonance (800 nm). a} The schematic illustration of the metasurface with the pump beams of 0.75 (blue), 1.5 (red), and 2.25 (yellow)$\mu$m in radius used in panel \textbf{b} and the pump intensities of 1.1 (deep blue), 2.2 (red), and 3.3 (green) mW/$\mu$m$^2$ in panel \textbf{f} with the radius of 1.5 $\mu$m. In all measurements, the probe size was 0.75 $\mu$m in radius. \textbf{b} The measured normalised $\Delta T/T$ of the metasurface at 800 nm probe for the 1.5 (blue), 3 (red), and 4.5 (yellow) $\mu$m 488 nm pump beams in diameter at 2.2 mW/$\mu$m$^2$. The inset shows the nonnormalised transmission,  $\Delta T/T$, of the same cases. \textbf{c} The simulated normalised $\Delta T/T$ of the 3 by 3 (blue), 7 by 7 (red), and 11 by 11 (yellow) array of nanodisks at 800 nm probe while pumping at 488 nm at 2 mW/$\mu$m$^2$. \textbf{d} The simulated transient (upper) and normalised transient (lower) temperatures of the 3 by 3 (blue), 7 by 7 (red), and 11 by 11 (yellow) array of nanodisks at 2 mW/$\mu$m$^2$ pump intensity. \textbf{e} The measured $\tau_{opt,heat}$ in \textbf{b} and calculated $\tau_{th,heat}$ in \textbf{d} versus pump diameter. \textbf{f} The measured normalised $\Delta T/T$ of the metasurface at 800 nm probe for 1.1 mW/$\mu$m$^2$ (deep blue), 2.2 mW/$\mu$m$^2$ (red), and 3.3 mW/$\mu$m$^2$ (green) intensity 488 nm 3 $\mu$m in diameter pump beams. The inset shows  $\Delta T/T$ of the same cases. \textbf{g} The simulated normalised $\Delta T/T$ of the 7 by 7 array of nanodisks for the 1 mW/$\mu$m$^2$ (deep blue), 2 mW/$\mu$m$^2$ (red), and 3 mW/$\mu$m$^2$ (green) 488 nm pump intensities at 800 nm probe wavelength. \textbf{h} The simulated transient (upper) and normalised transient (lower) temperatures of the 7 by 7 array of nanodisks for the 1 mW/$\mu$m$^2$ (deep blue), 2 mW/$\mu$m$^2$  (red), and 3 mW/$\mu$m$^2$ (green) 488 nm pump intensities. \textbf{i} The measured $\tau_{opt,heat}$ in \textbf{f} and calculated $\tau_{th,heat}$ in \textbf{h} versus pump intensity.
\\

\noindent \textbf{The measured and simulated transient TONL at near resonance probing. a} The transmission spectrum of the metasurface with ED resonance at 770 nm and the spectral positions of the pump (green) and probe (orange) wavelengths of 488 nm and 785 nm, respectively. \textbf{b} The measured normalised $\Delta T/T$ (upper) and nonrmalised temperature (lower) of the metasurface, for the 1.1 mW/$\mu$m$^2$ (blue), and 3.3 mW/$\mu$m$^2$ (yellow) 488 nm pump intensities at 785 nm probe wavelength. \textbf{c} The simulated normalised $\Delta T/T$ of the 7 by 7 array of nanodisks for the 1 mW/$\mu$m$^2$ (blue), and 3 mW/$\mu$m$^2$ (yellow) 488 nm pump intensities at 785 nm probe wavelength. \textbf{d} The measured pump intensity-dependent transmission of the metasurface at the pump and probe wavelengths of 488 and 785 nm, respectively. Upper x-axis shows the measured temperatures corresponding to the pump intensities. The blue (yellow) arrow illustrates the gradual change in transmission at 1.1 (3.3) mW/$\mu$m$^2$ in \textbf{b}. \textbf{e} The simulated $\Delta T/T$ at 785 nm (upper) for the laser is turned off at 300 ns at 3 mW/$\mu$m$^2$ pump intensity with 0.55 absorption at 488 nm and (lower) for the laser is turned off at 125 ns at 3 mW/$\mu$m$^2$ pump intensity with unity absorption at 488 nm.  \textbf{f} The simulated transient transmission at 785 nm for the modulated 488nm laser at 1 MHz with laser-on time equals 125 ns at 3 mW/$\mu$m$^2$ intensity with unity absorption at 488 nm. \textbf{g} The measured pump intensity-dependent transmission of the metasurface at the probe wavelength of 785 nm for the pump wavelengths of 488 nm (blue) and 785 nm (orange).

\bibliographystyle{unsrt}
\bibliography{main.bib} 
\end{document}